\documentclass[pra,twocolumn,showpacs,preprintnumbers,amsmath,amssymb,floatfix,superscriptaddress]{revtex4-1}

\usepackage{graphicx}
\usepackage{dcolumn}
\usepackage{bm,bbm}

\begin{document}


\title{Composite pulses for high-fidelity population inversion in optically dense, inhomogeneously 
broadened atomic ensembles}%

\author{Gabor Demeter}
\email{demeter.gabor@wigner.mta.hu}
\affiliation{Wigner Research Center for Physics, Hungarian Academy of Sciences, Konkoly-Thege
Mikl\'os \'ut 29-33, H-1121 Budapest, Hungary}

\date{\today}

\begin{abstract}
We derive composite pulse sequences that achieve high-fidelity excitation of two-state systems in 
an optically dense, inhomogeneously broadened ensemble. The composite pulses are resistant 
to distortions due to the back-action of the medium they propagate in and are able to 
create high-fidelity inversion to optical depths $\alpha z>10$. 
They function well with smooth pulse shapes used for  
coherent control of optical atomic transitions in quantum computation and communication. 
They are an intermediary solution 
between single $\pi$-pulse excitation schemes and adiabatic passage schemes, 
being far more error tolerant than the former but still considerably faster than the latter.

\end{abstract}

\maketitle

\section{Introduction}
\label{intro}

Coherent control methods that achieve the precise manipulation of atomic quantum states 
play a vital role in quantum information processing and quantum communication.
One class of methods that was applied extensively in 
this field utilizes various forms of adiabatic passage \cite{Vitanov2001,Kral2007}. 
While adiabatic methods possess robust fault tolerance, their slow speed can be a serious 
disadvantage.
Therefore alternative methods were developed using various optimization schemes with 
the aim of speeding the control process up, but at the same time retaining at least some of the 
fault tolerance \cite{Chen2010,Ruschhaupt2012,Torrontegui2013,Daems2013,Hegerfeldt2013}.
One of these alternative approaches is the method of composite pulses (also called composite pulse 
sequences) that was originally developed 
in NMR \cite{Levitt1986,Jones2003}, but has recently found its 
way into coherent control and quantum information processing 
\cite{Cummins2003,Roos2004,Torosov2011a,Torosov2011,Genov2011,Kyoseva2013,Torosov2014,Torosov2015,
Genov2014} .
 
One specific task in quantum information processing is building an optical quantum memory - a 
device that can store and retrieve the quantum state of a single-photon light pulse 
\cite{Lvovsky2009}. This is an indispensable component of quantum repeaters, devices that allow 
long-range quantum communication \cite{Sangouard2007,Sangouard2011}, but of also a number of other 
quantum technologies \cite{Bussieres2013}.
Memory schemes based on inhomogeneously broadened atomic ensembles (such as rare-earth ion doped 
optical crystals) and some variant of the photon 
echo effect have been studied intensively in recent years \cite{Tittel2010,Sangouard2011}.
Some of these achieve the rephasing of atomic coherences that is necessary for the echo emission by 
inverting (a part of) the atomic ensemble with laser pulses 
\cite{Damon2011,McAuslan2011,Dajczgewand2014}. The difficulty is that the ensemble must be 
optically dense to absorb the signal, which means that it will also distort the control pulses that 
are meant to produce population inversion. 
Thus inverting an inhomogeneously broadened, optically dense ensemble of atoms
is not a simple task, yet it is an important skill to master if these photon-echo 
type memories are to reach maturity.

Using single Rabi $\pi$-pulses in an optically dense ensemble is problematic because   
the pulse is quickly rendered ineffective by the medium 
\cite{Ruggiero2009,Demeter2013}. Control pulses that utilize adiabatic passage are more convenient 
\cite{Zafarullah2007,Damon2011,Demeter2013,Demeter2014}, but depending on the maximum 
pulse amplitude possible (limited for example by the maximum available laser power or the 
damage threshold of the crystal that hosts the ensemble) 
they may be too time consuming. The method of composite pulses (CPs) involves constructing 
complex control pulses for quantum state 
manipulation from a sequence of simple ``elementary'' pulses. 
Free parameters in the construction can be used to obtain fault tolerance with respect 
to various errors of the constituent pulses, such as frequency offsets or amplitude errors.
Even ``universal'' CPs can be developed that tolerate arbitrary imperfections.
The method of CPs has recently been used to develop error tolerant, high-fidelity population 
inversion schemes using smooth pulse shapes that are encountered in the optical regime 
\cite{Torosov2011a,Torosov2011,Genov2014}. In terms of speed and error tolerance, CPs can be 
regarded as a compromise between the single Rabi $\pi$-pulse and adiabatic passage methods.

In this paper we investigate the use of CPs for robust, high fidelity 
population inversion in inhomogeneously broadened, optically dense atomic ensembles for quantum 
information processing purposes. Similar to \cite{Torosov2011a} we try to perform the inversion 
using a CP built from a sequence of $N$ monochromatic Rabi $\pi$-pulses with appropriately chosen 
phases. We seek sets of phases (phase sequences) that allow the CP to invert an   
extended region within the ensemble (in terms of spectral width and optical depth) despite  
pulse distortions due to propagation in the medium. The usual approach in deriving CPs
assumes that the error that the phase sequence must compensate 
arises due to some imperfection of the experimental parameters such that it is reproduced for 
each elementary pulse of the sequence.  
In our case however, the pulses are distorted while propagating, and they are not all distorted 
the same way because some pulses excite the atoms while some return them to the ground 
state.

The paper is divided as follows. In section \ref{sec_formalism} we describe the basic physical 
setting and the equations to be solved. In section \ref{sec_altamp} we generalize the method of 
\cite{Torosov2011a} to derive CPs that are resilient with respect to amplitude errors due 
to the back action of the optically dense medium on the pulses. The key point here is that we 
allow the amplitude of even and odd numbered pulses of the sequence to change differently.
In section \ref{sec_combined} we derive CPs where 
amplitude error compensation is combined with a compensation of the atomic 
resonance frequency offset from the inhomogeneously broadened line center.
Finally in section \ref{sec_simulation} we present numerical simulations results of the   
Maxwell-Bloch equations for resonant pulse propagation. We also investigate the applicability 
of ``universal'' CPs derived recently in \cite{Genov2014} for high-fidelity inversion in optically 
dense ensembles.

\section{Computing propagators in an optically dense medium}
\label{sec_formalism}

The physical setting we consider consists of an inhomogeneously broadened, spatially extended 
ensemble of two-level atomic systems. The number density $\mathcal{N}$ of the absorbers is uniform 
in 
space and the precise resonance frequency of the $j$-th atom is offset by $\Delta_j$ 
from the inhomogeneous line center $\omega_0$: $\Delta_j=\omega_j-\omega_0$. The 
distribution of atomic frequencies $g(\Delta)$ is constant 
in space and assumed to be sufficiently wide to be taken a constant $g_0$
in the spectral region of the laser fields. A one dimensional propagation of 
CPs is considered along the $z$ direction, where the ensemble is optically 
dense. At the entry $z=0$, the CP consist of $N=2n+1$  
monochromatic pulses tuned to resonance 
with the atomic line center $\omega_0$.
Homogeneous decay processes are neglected and the ensemble is assumed to be in the ground state 
initially.

In this setting, the effect of the CP on the $j$-th atom located at $z_j$ is given by the 
Schr\"{o}dinger equation for the atomic probability amplitudes 
$|\psi_j\rangle=\alpha_j|g\rangle+\beta_j|e\rangle$:
\begin{align}
  \partial_t\alpha_j&=\frac{i}{2}\Omega^*(z_j,t)\beta_j\nonumber\\
  \partial_t\beta_j&=\frac{i}{2}\Omega(z_j,t)\alpha_j-i\Delta_j\beta_j.
\label{sch1}
 \end{align}
The complex Rabi frequency $\Omega(z_j,t)=E(z_j,t)d/\hbar$ characterizes 
the atom-field coupling [$E(z_j,t)$ is the slowly varying electric field amplitude at $z_j$, $d$ 
is the dipole matrix element]. 
In deriving Eq. \ref{sch1} 
the central frequency of the absorption line $\omega_0$ has been separated and the RWA 
applied.  Provided $\Omega(z_j,t)$ is known,
one can integrate Eq. \ref{sch1} and describe the effect of 
the CP on the atoms from the initial time $t_i$ until the final $t_f$ in 
terms of the unitary propagator 
$[\alpha_j(t_f),\beta_j(t_f)]^T=\mathbf{U}_j[\alpha_j(t_i),\beta_j(t_i)]^T$, which can
be conveniently written in terms of the complex Cayley-Klein parameters as:
\begin{equation*}
 \mathbf{U}_j=\left(\begin{array}{cc}
             a(\Delta_j,z_j) & b(\Delta_j,z_j) \\ -b(\Delta_j,z_j)^* & a(\Delta_j,z_j)^*
            \end{array}\right).
\end{equation*}
Clearly, the parameters $a(\Delta_j,z_j),
b(\Delta_j,z_j)$ depend on the location and frequency offset of the atom.
The problem is that the field is initially known only at the boundary and has to 
be computed for $z>0$ from the Maxwell 
equation for wave propagation. Using the slowly varying envelope 
approximation this can be written as:
\begin{equation}
 \left(\frac{1}{c}\partial_t+\partial_z\right)\Omega(z,t)=i\frac{\alpha}{\pi 
g_0}\mathcal{P}(z,t)
\label{maxwell}
\end{equation}
where $\alpha$ (without any subscript) is the absorption constant 
$\alpha=\pi g_0 k\mathcal{N}d^2/\varepsilon_0\hbar$ and the 
medium polarization $\mathcal{P}$  that constitutes the back-action of the ensemble on the 
propagating  
field is obtained by summing the atomic coherences within an infinitesimally thin region of $z$:
\begin{equation*}
 \mathcal{P}(z,t)=\sum_{j:z_j\in[z-dz,z+dz]} \alpha_j^*(t)\beta_j(t).
\end{equation*}

Equation \ref{maxwell}, and consequently the propagator $\mathbf{U}$ can usually be computed only 
numerically.
Furthermore, any result will pertain only to the specific initial state 
that the ensemble was in before the CP arrived. 
The choice we made (all atoms in $|g\rangle$) is 
adapted to quantum memory applications
where the absorption of a single- or few-photon signal pulse before the CP amounts to a 
negligible change in the ensemble initial state when the propagation of 
a strong classical control field is concerned. 
In accordance with the requirements encountered in several photon-echo type quantum memory schemes, 
we seek to establish high-fidelity atomic inversion in an extended region of the ensemble.
The figure of merit we use is the error probability that 
the atoms remain unexcited by the CP: $P_{err}=|\mathbf{U}_{11}|^2=|a(\Delta,z)|^2$, which is 
required to be as low as $10^{-2}-10^{-4}$ for high-fidelity quantum information applications.
The region must be wide enough around $\Delta=0$ to encompass the spectral width of the absorbed 
signal and extend to an optical depth of $\alpha z=5-10$. This latter requirement is defined by the 
fact that the signal, absorbed in the medium as $\sim\exp(-\alpha z)$ is contained in the region
$\alpha z<5$ up to an accuracy of $10^{-2}$ and in the region $\alpha z<10$ up to an accuracy of 
$10^{-4}$.

The standard procedure would now be to build the propagator of the CP from those of the constituent
``elementary'' pulses and use the free parameters to tune its effect on the atoms.   
In case the figure of merit used cannot 
be computed analytically with symbolic values of the parameters, a numerical optimization procedure 
can also be employed \cite{Roos2004}.
In our case however, not only is it impossible to obtain analytical 
formulas for the propagating fields with symbolic parameters, but the numerical solution of 
\ref{maxwell} is also expensive enough to make the use of numerical optimization practically 
impossible. 
Therefore we will use intuition to determine some conditions that we expect will improve the 
ability of the CP to invert extended regions in the optically dense, inhomogeneously broadened 
ensemble and then verify a posteriori that this is indeed the case. 
Similar to \cite{Torosov2011a}, we will assume that the CP is made up of $N=2n+1$ consecutive Rabi 
$\pi$-pulses, the only difference between them being an initial phase $\varphi_k$. It is also 
convenient to adapt the ``anagram'' condition $\varphi_k=\varphi_{N+1-k}$ and fix the overall phase 
of the CP by $\varphi_1=\varphi_N=0$. Thus the CP is characterized by the composite phase 
sequence $[0,\varphi_2,\varphi_3,\dots,\varphi_{n+1},\varphi_n,\dots,\varphi_2,0]$, defined 
entirely by
the set of $n$ phases $\{\varphi_k\}_{k=2}^{n+1}$ with which we can optimize the effect of our CP. 
We assume that the maximum Rabi frequency for all pulses is $\Omega_0$ (fixed either by 
available laser power or the damage threshold of the optical crystal that hosts the impurity ions) 
and compare the performance of all CPs to that of a single $\pi$-pulse.

\section{Amplitude error compensated composite pulses}
\label{sec_altamp}

For $\Delta=0$ Eqs. \ref{sch1} can be solved exactly to obtain the propagator for a single pulse
\begin{equation}
 \mathbf{U}_k(\varphi_k)=\left(\begin{array}{cc}
          \cos\frac{\mathcal{A}_k}{2} & i\sin\frac{\mathcal{A}_k}{2}e^{i\varphi_k}\\
          i\sin\frac{\mathcal{A}_k}{2}e^{-i\varphi_k} & \cos\frac{\mathcal{A}_k}{2}
         \end{array}\right)
\label{Uk}
\end{equation}
where the final level of atomic excitation is determined solely by 
the pulse area $\mathcal{A}=\int\Omega(t)dt$.
An elegant approach was developed in \cite{Torosov2011a} to derive 
CPs that are fault tolerant with respect to errors of pulse amplitude based on this fact,
which we will adapt to our problem.
The matrix \ref{Uk} was used to build the overall propagator as
\begin{equation}
\mathbf{U}=\mathbf{U}_N(0)\mathbf{U}_{N-1}(\varphi_2)
\dots\mathbf{U}_3(\varphi_3)\mathbf{U}_2(\varphi_2)\mathbf{U}_1(0)
\label{Useq}
\end{equation}
and, assuming that the amplitude of the constituent pulses was not perfect, 
$\mathcal{A}_k=\pi+\varepsilon$ was inserted for the imperfect pulse area. 
Then $a(\varepsilon)|_{\varepsilon=0}=0$ and various derivatives 
$\partial_\varepsilon^l a(\varepsilon)$ were nullified  
with appropriate choices of $\{\varphi_k\}$ 
to obtain CPs with considerable robustness against amplitude errors, having 
$P_{err}=\mathcal{O}(\varepsilon^{2N})$. (Since all even order derivatives disappear due to the 
anagram relation $\varphi_k=\varphi_{N+1-k}$, only odd order derivatives pose constraints for the 
phases.) All of the elementary pulses were 
identical in this approach (the error is duplicated identically for each one) and the result 
valid for arbitrary pulse shapes.

\begin{figure}[htb]
\includegraphics[width=0.48\textwidth]{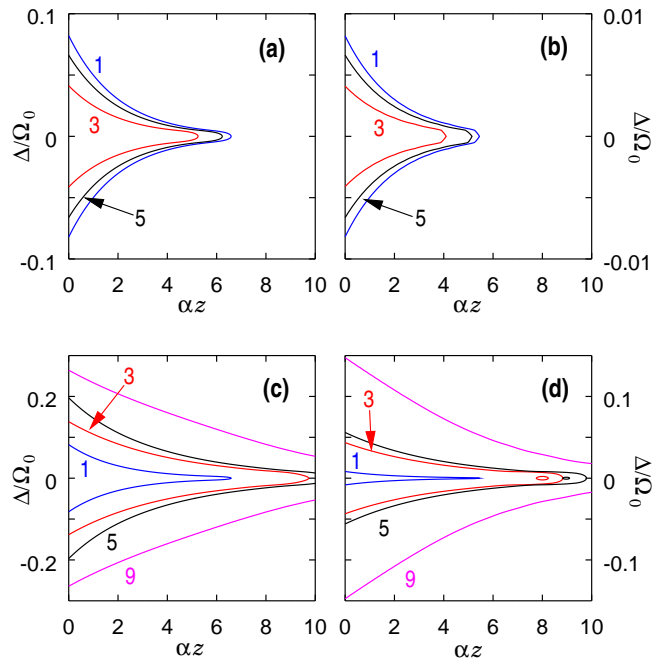}
\caption{Contour lines of the error probability $P_{err}=|a(\Delta,z)|^2$ for amplitude error 
compensated CPs of 
\cite{Torosov2011a} [a) and b)] and the alternating amplitude error compensated CPs
[ c) and d)] on the $\alpha z-\Delta$ parameter plane. a) $P_{err}=10^{-2}$ and b) $P_{err}=10^{-4}$
for a single pulse (blue line), the $N=3$ CP defined by $\{\varphi_2=2\pi/3\}$ (red line) and 
the $N=5$ CP with $\{\varphi_2=2\pi/5,\varphi_3=4\pi/5\}$ (black line). c) 
$P_{err}=10^{-2}$ and d) $P_{err}=10^{-4}$ for a single pulse (blue line), the $N=3$ CP designated
$U3a$ and defined by $\{\varphi_2=\pi/3\}$ (red line), the $N=5$ CP
$U5a_2:\{\varphi_2=\pi/5,\varphi_3=8\pi/5\}$ (black line) and the $N=9$ CP 
$U9a_8:\{\varphi_2=0.2708\pi,\varphi_3=1.0829\pi,\varphi_4=0.5898\pi,\varphi_5=4\pi/9\}$ 
(magenta line). Note that the scale on the $z$ axis is the same for all plots, but 
the scale on the $\Delta$ axis varies. The lines are tagged by $N$ on each subfigure. The first 
three phases for the 9 pulse sequence are 
approximate values.}
\label{fig_amp1}       
\end{figure}

Fault tolerance with respect to amplitude errors seems useful also when trying to invert atoms 
in an optically dense sample. It follows from the famed area theorem 
\cite{McCall1967,McCall1969} 
that for a single pulse, $\mathcal{A}=\pi$ is an unstable fixed point 
of the area equation, so any error will increase during propagation. Furthermore, even though a 
single perfect $\pi$-pulse should retain its area, due to energy loss it 
reshapes, gradually becoming longer - 
which means that its area will eventually start decreasing during the finite time interval 
allocated for the 
control. However, inserting the amplitude 
error compensated CPs of \cite{Torosov2011a} into the propagation equations shows that 
they do not perform better than single $\pi$-pulses at all. The 
results are shown in Fig. \ref{fig_amp1} a) and b) where the error contours $P_{err}=10^{-2}$ and 
$P_{err}=10^{-4}$ are plotted on the $\alpha z-\Delta$ plane (i.e. the boundary of 
the domain around 
$\Delta=0,z=0$ within which $P_{err}\leq 10^{-2},10^{-4}$). The data for the plot was obtained by 
computing $\mathbf{U}$ for a single incident $\cos^2$ shaped $\pi$-pulse, the amplitude error 
compensated $N=3$ CP defined by 
$\{\varphi_2=2\pi/3\}$ and the $N=5$ CP with 
$\{\varphi_2=2\pi/5,\varphi_3=4\pi/5\}$ \cite{Torosov2011a}. It is 
evident that while the single pulse can only produce high fidelity population inversion in an 
extremely limited domain of the ensemble, the $N=3$ and $N=5$ amplitude error compensated CPs are 
no 
better.

Now it also follows from the area theorem that for a pulse traveling in an 
inverted medium, the stability properties are reversed: 
$\mathcal{A}=\pi$ is the stable and $\mathcal{A}=2\pi$ is the unstable solution. 
Thus the area equation suggests the amplitude error that develops 
during propagation is not the same for all pulses, so assuming them to be 
identical 
is not justified in our case. Indeed it is intuitively clear that a pulse that must excite the 
atoms of the ensemble will be affected differently during propagation than one that returns them to 
the ground state. We thus write the error that we 
seek tolerance against in the following form:  
\begin{equation}
 \mathcal{A}_k=\pi+(-1)^k\varepsilon.
\label{amperror}
\end{equation}
This is perhaps the simplest expression that, without introducing new parameters, 
allows us to differentiate between odd numbered pulses of the composite sequence 
(i.e. those that are to excite the atoms) and even numbered ones (those which are to return them 
to the ground state).
We will refer to this ansatz as alternating amplitude error. 
Inserting \ref{amperror} into Eq. \ref{Uk}, composing the overall
propagator \ref{Useq} and using the constraints 
$\partial_\varepsilon^l a(\varepsilon)=0$ for various sets of 
derivatives, we can derive CPs that are fault tolerant with respect
to alternating amplitude errors. For $N=3$, the simplest case, the condition  
$\partial_\varepsilon a=0$ gives $\cos\varphi_2=1/2$ which is solved by  
$\varphi_2=\pi/3$, defining the CP designated by $U3a$.  Another example 
is the $N=5$ case where the conditions  
$\partial_\varepsilon a=0$ and $\partial_\varepsilon^3 a=0$ 
eventually yield $4\cos\varphi_2+2\cos(2\varphi_2-\varphi_3)-2\cos\varphi_3=1$ and
$\cos(\varphi_2-\varphi_3)-\cos(2\varphi_2-\varphi_3)=1/2$.
($\partial_\varepsilon^2 a=0$ and  $\partial_\varepsilon^4 a=0$ are 
satisfied automatically.) Solving these equations we obtain the 
sequences $U5a_1:\{\varphi_2=3\pi/5,\varphi_3=4\pi/5\}$ and
$U5a_2:\{\varphi_2=\pi/5,\varphi_3=8\pi/5\}$.
(The designation $U5a_j$ means that it is the $j$-th phase sequence with $N=5$ and alternating 
amplitude error compensation.)
Since all even order derivatives of $a(\varepsilon)$ 
disappear due to the anagram relation, for an $N$ pulse CP where we have $n=(N-1)/2$ 
phases to nullify derivatives, the first nonzero derivative will be the $N$-th order one, so 
$P_{err}=\mathcal{O}(\varepsilon^{2N})$. The procedure can be continued to higher orders, but of 
course the resulting trigonometric equations will be progressively more difficult to solve.
Up to $N=9$ we have found, for the $N=2n+1$ phase sequence, $2^{n-1}$ solutions. 
All the phases obtained are tabulated in table \ref{table_diffamp} of the appendix.

Inserting the CPs derived in this manner in Eq. \ref{maxwell} we can verify 
that they are indeed more fault tolerant with respect to propagation induced
distortions than a single $\pi$-pulse. Figure \ref{fig_amp1} c) and 
d) show the $P_{err}=10^{-2}$ and $P_{err}=10^{-4}$ contours for several CPs together 
with the single pulse case. It can be seen that increasing $N$ leads to the expansion of the 
high-fidelity population transfer domain. For $N=9$ the $P_{err}=10^{-4}$ domain 
already 
extends past $\alpha z=10$ and it is also about an order of magnitude wider in $\Delta$ than for 
the single pulse case. The performance of a CP can 
only be evaluated by computing the propagating fields numerically - the 
CPs depicted are the ones that perform best for each $N$. 
An $N=7$ sequence was omitted in the figure because the best 7 pulse CP was only slightly better 
than the $N=5$ one depicted. Note that if the sequence $\{\varphi_k\}$ is a valid 
solution, then so is $\{-\varphi_k\}$ and thus $\{2\pi-\varphi_k\}$ - we consider
these sequences to be identical (they are not listed in table \ref{table_diffamp}). 
The figures were produced by 
using $\cos^2$ shaped pulses - they are convenient to use because they become exactly zero at a 
finite time point.

\section{Frequency offset and amplitude error compensation}
\label{sec_combined}

Composite pulses with alternating amplitude error compensation allow the $P_{err}<10^{-4}$ region 
to reach $\alpha z>10$, which is perfectly sufficient. The region is still fairly narrow with 
respect to the frequency offset $\Delta$, so we now seek to derive CPs with combined error 
compensation of alternating amplitude error and frequency offset. However, when $\Delta\neq0$ the 
Cayley-Klein parameters depend on the shape of the pulse envelope, so we turn to pulse 
shapes for which Eqs. \ref{sch1} are analytically solvable.    

\subsection{Hyperbolic-secant pulse}
\label{sec_hypsec}

First we use the following solution for the Rosen-Zener model \cite{Rosen1932,Torosov2011a}, for 
which $\Omega(t)=\Omega_0\mathrm{sech}(t/T)$ and the frequency offset (detuning) is constant. The 
Cayley-Klein parameters for the $k$-th pulse with phase $\varphi_k$ are:
\begin{align}
a_k&=\frac{\Gamma\left(\frac{1}{2}+iq\right)^2}
{\Gamma\left(\frac{1}{2}+iq-p\right)\Gamma\left(\frac{1}{2}+iq+p\right)}\nonumber\\
b_k&=i\frac{\sin\pi p}{\cosh\pi q}e^{i\varphi_k}
\label{eq_sechparameters}
\end{align}
where $p=\Omega_0T/2$ and $q=\Delta T/2$. Clearly, for $\Omega_0T=1$ we have the $\pi$-pulse 
case, $b=ie^{i\varphi},a=0$. 
We insert the alternating amplitude error ansatz for the $m$-th pulse 
as $\Omega_0T=1+(-1)^m\varepsilon$ in 
Eqs. \ref{eq_sechparameters} and create the CP propagator via \ref{Useq} as before. We then seek 
phase sequences where various derivatives of $a(\varepsilon,\Delta)$ 
 are zero at $\varepsilon=0,\Delta=0$. 

For $N=3$ we have two equations from the first derivatives
$\partial_\Delta a=0$ and
$\partial_\varepsilon a=0$, but they both yield the same constraint for $\varphi_2$:
$\cos\varphi_2=1/2$ which is satisfied by $\varphi_2=\pi/3$. The CP designated $U3c$ and defined by 
$\{\varphi_2=\pi/3\}$ has been derived
in \cite{Torosov2011a} as a detuning compensated CP, and the previous section of the present paper 
as the alternating amplitude error compensated CP $U3a$ for arbitrary pulse shapes. We now see that 
for 
hyperbolic-secant pulse shape it is a CP with combined frequency offset / alternating amplitude 
error compensation. The order of the error for 
this sequence is $P_{err}=\mathcal{O}(\Delta^4),\mathcal{O}(\varepsilon^4)$. 

For $N=5$ the equations 
$\partial_\Delta a=0$ and
$\partial_\varepsilon a=0$ again both yield the same constraint for
the phases: 
\begin{equation}
1-2\cos(\varphi_2-\varphi_3)+2\cos(2\varphi_2-\varphi_3)=0.
\label{cond5a}
\end{equation}
Computing the second derivatives shows that the conditions 
obtained from the equations $\partial_\Delta^2a=0$ and
$\partial_\varepsilon\partial_\Delta a=0$ are also equivalent:
\begin{equation}
1+2\cos(\varphi_2-\varphi_3)+2\cos(2\varphi_2-\varphi_3)=0.
\label{cond5b}
\end{equation}
($\partial_\varepsilon^2a=0$ is 
automatically fulfilled).
These two conditions are then 
satisfied by two phase sequences: $U5c_1:\{\varphi_2=5\pi/6,\varphi_3=\pi/3\}$  and 
$U5c_2:\{\varphi_2=\pi/6,\varphi_3=5\pi/3\}$. (The designation of the form $U5c_i$ means that it is 
the $i$-th CP with $N=5$ and combined error 
compensation.) $P_{err}\sim\mathcal{O}(\Delta^6),\mathcal{O}(\varepsilon^6)$ for these sequences.

For $N=7$, $\partial_\Delta a=0$ and
$\partial_\varepsilon a=0$ again yield a single constraint
\begin{multline}
 1-2\cos(\varphi_3-\varphi_4)+2\cos(\varphi_2-2\varphi_3+\varphi_4)\\
-2\cos(2\varphi_2-2\varphi_3
+\varphi_4) = 0,
\label{cond7a}
\end{multline}
while  $\partial_\Delta^2 a=0$ and
$\partial_\varepsilon\partial_\Delta a=0$ both reduce to:
\begin{multline}
1+2\cos(\varphi_3-\varphi_4)+2\cos(\varphi_2-2\varphi_3+\varphi_4)\\
+2\cos(2\varphi_2-2\varphi_3
+\varphi_4)=0.
\label{cond7b} 
\end{multline}
Computing all third order derivatives we find that from $\partial_\Delta^3 a=0$, 
$\partial_\varepsilon\partial_\Delta^2 a=0$,
$\partial_\varepsilon^2\partial_\Delta a=0$ and
$\partial_\varepsilon^3 a=0$ we 
obtain only one more constraint for the phases.
Any one of these four equations can be used together with Eqs. \ref{cond7a} and \ref{cond7b}
to obtain the same phase sequences because any of the third order derivatives can be
expressed as a linear combination of another one, (say $\partial_\Delta^3 a$) and Eqs. \ref{cond7a},
\ref{cond7b}. Thus for $N=7$ we can nullify all 
derivatives up to third order using 
six phase sequences tabulated in table \ref{table_comb} of the appendix as $U7c_1-U7c_6$.
These sequences yield CPs with $P_{err}=\mathcal{O}(\Delta^8),\mathcal{O}(\varepsilon^8)$.

Finally we have considered the $N=9$ case. Similar to the $N=3,5$ and $N=7$ cases, we found that
all the equations from a given order of derivatives essentially yield only one additional 
independent constraint for the phases. Thus solving four trigonometric equations gives us phase 
sequences 
where all derivatives of $a(\varepsilon,\Delta)$ will be nullified up to fourth order, yielding CPs
with $P_{err}\sim\mathcal{O}(\varepsilon^{10}),\mathcal{O}(\Delta^{10})$. The twelve phase 
sequences obtained are tabulated in the appendix in table \ref{table_comb} for reference, designated
$U9c_1-U9c_{12}$.

It is remarkable, that for hyperbolic-secant pulses, we have been able to nullify all the 
derivatives up to $n$-th order with $n$ phase parameters all the way up to $n=4$ (i.e. $N=9$).
For $N=11$ and higher the expressions for the derivatives become intractable, so it is unclear 
whether this trend continues. However it must be noted that nullifying so many derivatives with so 
few parameters is possible because of the symmetries inherent in the alternating amplitude error 
ansatz \ref{amperror} and the anagram relation together. A less symmetric amplitude error ansatz 
can easily lead to constraints that can not all be satisfied simultaneously. Relaxing the anagram 
relation allows more phase parameters for the same pulse number $N$, but more independent 
constraints per level of derivatives - overall there is no gain.

\subsection{Comments on other pulse shapes}

Two important questions immediately arise. First, do the CP-s  
derived in the previous subsection actually show improved performance 
when propagating in the ensemble and how does this performance increase with 
the number of pulses $N$? Second, are the results applicable to other pulse shapes?
The second question is 
important because hyperbolic-secant pulses fall off very slowly, so, 
with the time required for the operation being an important factor, more compact pulses such 
as Gaussian or $\cos^2$ pulse shapes are preferable. 

To gain insight into the second question, we have repeated the derivation outlined above for 
square pulses. While not very relevant for short pulses in the optical domain, they too allow 
analytic solution of Eqs. \ref{sch1}. 
We have found, that up to $N=7$ we obtain precisely the same constraints and solutions for the 
phases as for hyperbolic-secant pulses.
However, for $N=9$ the square-pulse case deviates. The equations obtained from the 
fourth order derivatives can no longer be reduced to a single constraint.
Thus the four phases we have for $N=9$ are no longer sufficient to nullify 
all fourth order derivatives for square pulses. 

More insight can be gained by noting that the 3 and 5 pulse CPs have been derived before.
First, $U3a/U3c$, has been derived in \cite{Torosov2011a} as a detuning compensated 
sequence and it has also been stated that up to $N=5$ detuning compensated CPs are 
independent of the pulse shape, provided that it is symmetric in time: $\Omega(t)=\Omega(-t)$.
Next, $U5c_1$ has also been found in \cite{Torosov2011a} as a CP with simultaneous 
compensation of amplitude error and detuning. Finally, $U5c_1$ and $U5c_2$ have been derived in 
\cite{Genov2014} 
(denoted $U5a$ and $U5b$ there) as ``universal'' CPs, sequences that compensate pulse errors to 
first order regardless of their nature (imperfect pulse shape, amplitude, detuning, chirp, etc. - 
provided that it is exactly reproduced for all of the pulses).

Repeating the derivation in subsection \ref{sec_hypsec} using uniform amplitude error 
$\Omega_0T=1+\varepsilon$, we find that for $N=5$,
$\partial_\Delta a=0$ yields condition \ref{cond5a}, while $\partial_\varepsilon a =0$ reduces to 
condition \ref{cond5b}. Of the second order derivatives, $\partial_\Delta^2 a=0$ yields 
constraint \ref{cond5b}
and $\partial_\Delta\partial_\varepsilon a=0$ yields constraint \ref{cond5a}. 
So for the uniform amplitude error considered in \cite{Torosov2011a} we need two phases to nullify 
both 
first order 
derivatives - but then the second order ones are simultaneously nullified as well.
Thus $U5c_1$ and $U5c_2$ are in fact CPs that compensate both frequency offset (detuning) and 
uniform amplitude error / alternating amplitude error simultaneously up to second order. 
All these point to $U5c_1$ and $U5c_2$ being combined 
alternating amplitude error / frequency offset compensated sequences for arbitrary 
(symmetric) pulse shapes.

\section{Simulation results for CP propagation}
\label{sec_simulation}

To verify that the CPs derived in section \ref{sec_hypsec} do perform better at inverting an 
optically dense ensemble, 
we have computed their propagation for three different smooth pulse envelopes using Eq. 
\ref{maxwell} - hyperbolic-secant: $\Omega_s(t)=\Omega_0\mathrm{sech}(t/T_s)$, 
Gaussian: $\Omega_g(t)=\Omega_0\exp(-t^2/2T_g^2)$ and $\cos^2$:
$\Omega_c(t)=\Omega_0\cos^2(t\pi/2T_c)$. The same peak pulse amplitude $\Omega_0$ was used 
in all three cases and the time constants were adjusted so 
that pulse areas were $\pi$: $T_s=1$, $T_g=\sqrt{\pi/2}$ and $T_c=\pi$. Using the field 
$\Omega(z,t)$ obtained from the simulation,  
$\mathbf{U}(\Delta,z)$ was computed and the error contours $P_{err}=|a(\Delta,z)|^2$ plotted on 
the $\alpha z-\Delta$ plane. 

\begin{figure}[htb]
\includegraphics[width=0.45\textwidth]{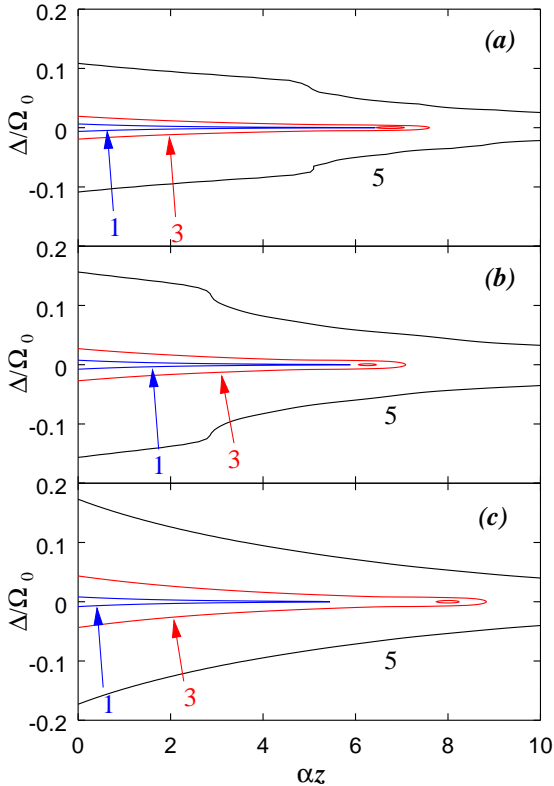}
\caption{Contour lines of $P_{err}=10^{-4}$ on the ${\alpha}z \mathrm-\Delta$ plane 
for a single $\pi$-pulse (line 1), the $N=3$ CP $U3c:\{\varphi_2=\pi/3\}$ (line 3) and the $N=5$ 
CP $U5c_2:\{\varphi_2=\pi/6,\varphi_3=5\pi/3\}$ (line 5) in case of a) hyperbolic-secant 
pulses, 
b) Gaussian pulses and c) $\cos^2$ shaped pulses.}
\label{fig_comb2}       
\end{figure}

Figure \ref{fig_comb2} shows the $P_{err}=10^{-4}$ contours for a) hyperbolic-secant, b) Gaussian 
and c) $\cos^2$ pulse shapes. The lines tagged by 1, 3 and 5 belong to a single $\pi$-pulse, the 
 $U3c$ and the $U5c_2$ sequences respectively on all three subfigures. One can see that 
for all three pulse 
shapes the $U3c$ and $U5c_2$ sequences show greatly improved performance over the single
$\pi$-pulse. The spectral width of the high-fidelity population transfer region for the $U5c_2$ CP 
is about 20 times wider at $\alpha z=0$ and about 120 times wider at $\alpha z=5$ for all three 
pulse shapes. This corroborates the conjecture that these sequences are universal and should work 
for 
arbitrary symmetric pulse shapes. It is important to note that the two $N=5$ CPs do 
not perform equally well. $U5c_1$ (not depicted here) is considerably better than the 
$U3c$ sequence, but is systematically inferior to the $U5c_2$ one. This is because 
after all second order derivatives are nullified, the (nonzero) values of the third order 
derivatives will matter most. For the hyperbolic-secant pulse where these 
derivatives can be evaluated, we found the four third order 
derivatives to be about an order of magnitude smaller for the $U5c_2$ sequence.

\begin{figure}[htb]
\includegraphics[width=0.45\textwidth]{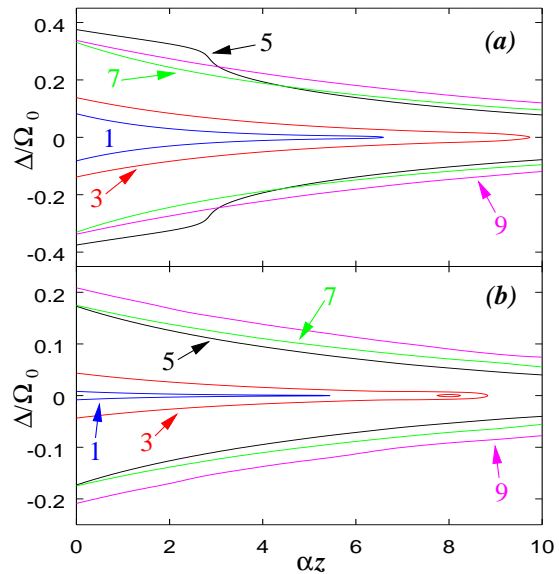}
\caption{Contour lines of a) $P_{err}=10^{-2}$ and b) $P_{err}=10^{-4}$ on the ${\alpha}z 
\mathrm-\Delta$ plane. The lines tagged by 1, 3, 5, 7 and 9 correspond to the single-pulse case and 
to the CPs $U3c:\{\varphi_2=\pi/3\}$, 
$U5c_2:\{\varphi_2=\pi/6,\varphi_3=5\pi/3\}$, 
$U7c_1:\{\varphi_2=-0.647\pi,\varphi_3=\pi/3,\varphi_4=0.647\pi\}$ and
$U9c_1:\{\varphi_2=0.025\pi,\varphi_3=0.847\pi,\varphi_4=0.670\pi,\varphi_5=1.299\pi\}$ 
respectively.
Phases in fractional form are exact values while phases in decimal form are approximate.
}
\label{fig_comb1}       
\end{figure}

Because the $\cos^2$ pulses are the most compact of the above three, (they become exactly zero 
at finite $t$), this is the pulse shape we have used for 
investigating longer sequences. Figure 
\ref{fig_comb1} shows the $P_{err}=10^{-2}$ and the $P_{err}=10^{-4}$ contours for several CPs
up to $N=9$ on the $\alpha z-\Delta$ plane. Evidently the performance does improve with 
$N$, but the improvement slows considerably after $N=5$. This however is not surprising, 
as the phases were derived using a series expansion. One can also see that for lower fidelity 
and optical depth, $U5c_2$ beats even the 7 and 9 pulse CPs depicted.
Note that we have plotted, out of the six CPs designated $U7c_1-U7c_6$ and of the 
12 designated $U9c_1-U9c_{12}$, the ones 
that show the best performance. It is not 
strictly true that each $N=9$ CP is better than any $N=7$ CP. 
This effect is probably due to the values of 
higher order derivatives that are not nullified as we have seen for $U5c_1$ and $U5c_2$.
Overall, even though the duration of a $N=9$ pulse CP is considerably shorter than pulses that 
would achieve the 
population transfer via AP, the best tradeoff seems to be $N=5$, where the $P_{err}<10^{-4}$ region
already reaches $\alpha z=10$. Note also that CPs with combined error compensation are superior to
CPs with solely alternating amplitude error compensation. Comparing the lines on Fig. 
\ref{fig_amp1} d) and \ref{fig_comb1} c) (both represent $\cos^2$ shaped pulses) one can see that 
the $N=5$ CP with combined error compensation is better than the $N=9$ CP with only amplitude error 
compensation.

As the sequences $U5c_1$ and $U5c_2$ derived here have also turned up as ``universal composite 
pulses'' in a very
different setting in \cite{Genov2014}, we carried out a detailed investigation
of all the CPs derived there as they propagate in the optically dense medium.
The derivation there did not
assume any constraint on the nature of the compensated imperfection or the shape of the pulse 
(hence the term ``universal''), only that it is exactly the same for all elementary pulses. 
Phase sequences were derived for $N=5,7$ and $9$ pulse CPs which compensate the errors up to second 
order, $N=13$ CPs that compensate them to fourth order and $N=25$ CPs that do so to eighth order.

We have substituted the phases tabulated in table I. of \cite{Genov2014} for the CPs
$U7a$, $U7b$, $U9a$ and $U9b$ into the constraints derived 
in section \ref{sec_hypsec} for hyperbolic-secant pulses. We found the sequences to satisfy the 
constraints derived from the first and second order derivatives, but not the third order ones.
Correspondingly, computing their propagation using Eqs. \ref{maxwell} we found that 
their performance is practically the same as that of the $U5c_2$ CP (the shortest one that 
allows combined error compensation to second order). By comparison, the $U7c_1$ and $U9c_1$ 
CPs derived in the present work did function better than $U5c_2$, even if the increase in 
performance was not as prominent as during the $N=3$ to $N=5$ step.

\begin{figure}[htb]
\includegraphics[width=0.5\textwidth]{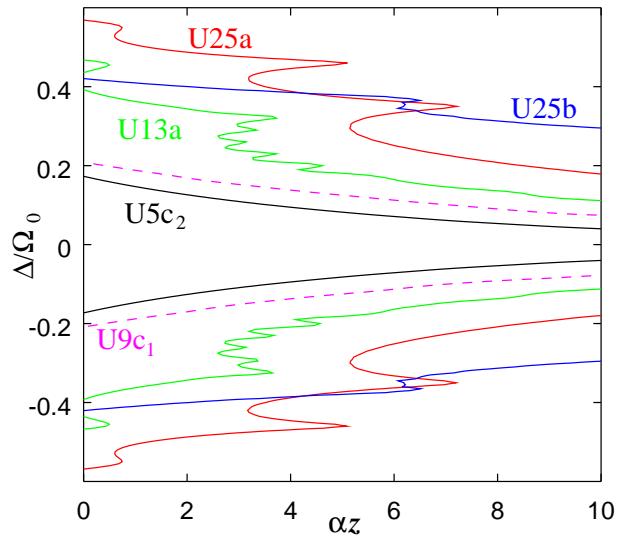}
\caption{Contour lines of $P_{err}=10^{-4}$ on the ${\alpha}z 
\mathrm-\Delta$ plane for some CPs derived in \cite{Genov2014}.  The lines tagged by $U5c_2$ 
(black) and $U9c_1$ (magenta - dashed line) correspond to CPs derived in this paper (as well).
}
\label{fig_universal}       
\end{figure}

While we have not been able to derive phases for combined error compensation CPs higher than $N=9$, 
it is still interesting to check the performance of the  $N=13$ and $N=25$ ``universal'' sequences.
Figure \ref{fig_universal} shows the error contours that mark the boundary of high-fidelity 
inversion for several CPs. For $N=13$, we have plotted $U13a$ which is clearly better, 
but for $N=25$ the two CPs $U25a$ and $U25b$ are both shown - the former is better for smaller 
optical depths while the latter is superior for larger ones. For a comparison, the  
$U9c_1$ CP derived here and the $U5c_2$ CP derived in both papers are also shown. 
Clearly, there is a considerable increase in 
performance with $N$ which is intriguing because it suggests that CPs with $N>9$ 
and combined alternating amplitude / frequency offset error compensation can be derived to 
achieve still higher performance in inverting an optically dense ensemble.  
One must keep in mind however, that very long sequences (such as $N=25$) are not practical because
the time required already allows adiabatic passage schemes to function better. 
 
Another problem must be considered when employing CPs for inverting optically dense ensembles, 
namely the time window in which the amplitude of the pulse sequence is appreciably different from 
zero at a given optical depth. A single $\pi$-pulse increases its temporal length as it propagates, 
developing a long ``tail'' that can overlap any signal echos that are to be retrieved from the 
ensemble \cite{Ruggiero2009}, making its application impractical. A similar effect can be observed
with some of the CPs studied in this paper. For some phase sequences a long oscillatory tail 
develops as the CP propagates, increasing the time window where the control field 
is non-negligible. This does not affect all sequences equally, some CPs are much less affected 
than others. It is a property that has to be investigated for each sequence separately 
when considering its application in photon-echo quantum memory schemes. 

Finally, some comments on the validity of the two-level model without relaxation are in order. To 
achieve high-fidelity quantum state control, the entire manipulation process has to be concluded in 
a time $T_{CP}$ much less than the lifetime $T_1$ of the atomic excited state (or the coherence 
lifetime $T_2$ if that is shorter). Given the requirement $P_{err}<10^{-2},10^{-4}$, we can readily 
estimate the time available as $T_{CP}<P_{err}T_1$. (A 
more precise calculation using a master equation for the two-state system shows that this is in 
fact an overestimation, but it is very useful for order of magnitude considerations.)
Conversely, if the interaction time is shorter than this limit, relaxation can be neglected for our 
purposes. For a 
number of rare-earth ions such as erbium, thulium, europium or terbium that can be described as 
two-state systems in optical crystals under certain conditions, excited state lifetimes in the 
range of 1-10 ms were measured \cite{Thiel2011}.
The time available for the manipulation in these cases is then $10-100\mu s$ (for 
$P_{err}=10^{-2}$) and $0.1-1\mu s$ (for $P_{err}=10^{-4}$). 
Of course, for longer composite sequences this means shorter elementary 
pulses, so very long sequences are inconvenient.

\section{Summary and Outlook}

In this paper, we have investigated the use of composite pulses for the high-fidelity inversion of 
two-level systems in an optically dense, inhomogeneously broadened ensemble. Such ensembles, found 
for example in rare-earth doped optical crystals, have important applications in quantum 
communication and quantum computing , e.g. as a medium for realizing photon-echo based optical 
quantum memories. High-fidelity inversion in optically dense media is problematic because 
they distort the pulses as they propagate.

We have introduced the concept of 
alternating amplitude error (amplitude error that 
is of opposite sign for even and odd numbered pulses of the sequence)
and have derived phase sequences that grant the CP robustness with respect to it.
We have shown that these CPs are then able to invert the atoms of the ensemble to a greater optical 
depth than single $\pi$-pulses.
When alternating amplitude error compensation is combined with 
frequency offset compensation, we obtain CPs that are even more effective.
Using CPs made up of of as few as five pulses, the region of high-fidelity inversion can easily 
reach an optical depth of $\alpha z=10$ and be over two orders of magnitude wider in spectral width 
than the inversion obtained using a single $\pi$-pulse. 
The phase sequences were derived using series expansions of various analytically solvable 
models for two-level atomic excitation. Their performance in creating high-fidelity inversion
within the ensemble was then verified by numerical simulation of the Maxwell-Bloch equations
for pulse propagation.   
Finally, 
we have also verified that some of the universal composite pulses derived in \cite{Genov2014} 
are also effective at creating high-fidelity inversion in optically dense ensembles.

Overall, some of the CPs derived demonstrate great potential for inverting optically dense ensembles
- they are much more robust than single monochromatic $\pi$-pulses, but can be considerably faster 
than adiabatic passage methods. Furthermore, one can possibly refine further  
the phase sequences using numerical optimization schemes. Numerical schemes that need to
solve the Maxwell-Bloch equations at each step are far too expensive computationally to perform
an optimization from the start. They may, however be suitable to obtain an even higher 
performance by using one of the sets of phases derived in this paper as a starting point
and executing the optimization scheme only for a very limited number of steps.

\acknowledgments
The numerical computations of the paper were carried out with the use of  GNU Octave \cite{octave}.

\appendix*

\section{Complete list of phases}

Below is a list of phases for the alternating amplitude error compensated CPs derived in section 
\ref{sec_altamp} and the combined error compensation CPs from section \ref{sec_combined}.
 \setlength{\tabcolsep}{8pt}
\begin{table}[htb]
\begin{tabular}{c|c c c c}
 designation & $\varphi_2$ & $\varphi_3$ & $\varphi_4$ & $\varphi_5$\\
\hline
&&&&\\
$U3a$ & 1/3 &  &  & \\
$U5a_1$ & 3/5 & 4/5  &  & \\
$U5a_2$ & 1/5 & 8/5  &  & \\
$U7a_1$ & 1/7 & 10/7 & 13/7 & \\
$U7a_2$ & 0.230 & 1.230 & 1 &  \\
$U7a_3$ & 3/7 & 2/7 & 11/7 & \\
$U7a_4$ & 5/7 & 8/7 & 9/7 & \\
$U9a_1$ & 1/9 & 4/3 & 5/3 & 10/9 \\
$U9a_2$ & 1/3 & 0 & 1/3 & 2/3\\
$U9a_3$ & 1/3 & 0 & 5/3 & 0\\
$U9a_4$ & 5/9 & 2/3 & 1/3 & 14/9\\
$U9a_5$ & 7/9 & 4/3 & 5/3 & 16/9\\
$U9a_6$ & 0.145 & 1.280 & 0.024 & 2/9\\
$U9a_7$ & 0.199 & 1.803 & 1.160 & 8/9\\
$U9a_8$ & 0.271 & 1.083 & 0.590 & 4/9
\end{tabular}
\caption{List of phases given as multiples of $\pi$ for alternating error compensated CPs 
derived in section \ref{sec_altamp}. 
Phases in fractional form are exact values while phases in decimal form are approximate.}
\label{table_diffamp} 
\end{table}

\begin{table}[htb]
\begin{tabular}{c|c c c c}
 designation & $\varphi_2$ & $\varphi_3$ & $\varphi_4$ & $\varphi_5$\\
\hline
&&&&\\
$U3c$ & 1/3 &  &  & \\
$U5c_1$ & 5/6 & 1/3  &  & \\
$U5c_2$ & 1/6 & 5/3 &  &\\
$U7c_1$ & -0.647 & 1/3 & 0.647 &\\
$U7c_2$ & -0.176 & 1/3 & 0.176 & \\
$U7c_3$ & 0.425 & 1/3 & -0.425 &\\
$U7c_4$ & 0.536 & 1/3 & -0.536 & \\
$U7c_5$ & 0.193 & 1.386 & 1.245 &\\
$U7c_6$ & 0.955 & 0.911 & 1.533 & \\
$U9c_1$ & 0.025 & 0.847 & 0.670 & 1.299\\
$U9c_2$ & 0.057 & 1.30 & 1.911 & 0.095 \\
$U9c_3$ & 0.126 & 1.454 & 1.916 & 1.983\\
$U9c_4$ & 0.128 & 1.458 & 1.789 & 1.724\\
$U9c_5$ & 0.431 & 1.189 & 0.572 & 0.385\\
$U9c_6$ & 0.500 & 0.606 & 0.304 & 1.601\\
$U9c_7$ & 0.526 & 0.609 & 0.294 & 1.595\\
$U9c_8$ & 0.721 & 0.254 & 0.878 & 1.925 \\
$U9c_9$ & 0.731 & 0.261 & 0.791 & 1.721\\
$U9c_{10}$ & 0.782 & 1.474 & 0.530 & 0.212\\
$U9c_{11}$ & 0.808 & 0.779 & 1.751 & 0.084\\
$U9c_{12}$ & 0.866 & 0.570 & 0.814 & 1.585
\end{tabular}
\caption{List of phases given as multiples of $\pi$ for combined error compensated CPs derived in 
section \ref{sec_hypsec}. 
Phases in fractional form are exact values while phases in decimal form are approximate.}
\label{table_comb} 
\end{table}

\clearpage

\bibliography{/home/gdemeter/fiz/manuscript/bibliography}

\bibliographystyle{apsrev4-1}


\end{document}